\documentclass[twocolumn, showpacs,preprintnumbers,amsmath,amssymb]{revtex4-1}
\usepackage{amssymb}
\usepackage{xcolor}
\usepackage{graphicx}
\usepackage{dcolumn}
\usepackage{bm}
\usepackage{multirow}
\usepackage{booktabs}
\usepackage{natbib}
\usepackage{soul}
\usepackage{amssymb}
\usepackage{amsmath}	
\usepackage{color,soul}
\usepackage{cases} 
\usepackage{mhchem}
\usepackage{txfonts}

\newcommand{\foun}{\ce{^{14}N}}
\newcommand{\fifn}{\ce{^{15}N}}

\begin{document}
\preprint{APS/123-QED}

\title{Dissociative recombination of N$_2$H$^+$: Isotopic effects}

\author{J. Zs. Mezei$^{1,2}$}\email[]{mezei.zsolt@atomki.hu}
\author{A. Orb\'an$^{1}$}
\author{S. Demes$^{1}$}
\author{M. Ayouz$^{3,4}$}
\author{A. Faure$^{5}$}
\author{P. Hily-Blant$^{5}$}
\author{Ioan F. Schneider$^{2,6}$}
\affiliation{$^{1}$HUN-REN Institute for Nuclear Research (ATOMKI), H-4001 Debrecen, Hungary}%
\affiliation{$^{2}$LOMC CNRS, Universit\'e le Havre Normandie, F-76058 Le Havre, France}%
\affiliation{$^{3}$CentraleSup\'elec, Universit\'e Paris-Saclay, SPMS, Gif-sur-Yvette, F-91190, France}%
\affiliation{$^{4}$CentraleSup\'elec, Universit\'e Paris-Saclay, LGPM, Gif-sur-Yvette, F-91190, France}%
\affiliation{$^{5}$Univ. Grenoble Alpes, CNRS, IPAG, 38000 Grenoble, France}%
\affiliation{$^{6}$LAC, CNRS Universit\'e Paris-Saclay, F-91405 Orsay, France}%
\date{\today}

\begin{abstract}
   The investigation of the isotopic ratio of interstellar nitrogen  - $^{14}$N versus $^{15}$N - is done, for explaining its variations observed for N$_2$H$^+$ in different interstellar and Solar environments. 
     The goal is to produce cross sections and rate coefficients for electron impact dissociative recombination for different isotopologues of N$_2$H$^+$, since it was envisioned as a novel source that can lead to nitrogen fractionation.
   We calculate dissociative recombination cross sections and rate coefficients using the normal mode approximation combined with the R-matrix theory and vibronic frame transformation within the multichannel quantum defect theory for eight isotopologues containing both $^{14}$N, $^{15}$N and H, D.
   Our calculations show that the relative differences respective to the main isotopologue ($^{14}$N$_2$H$^+$) is below 1\% for the hydrogen containing isotopologues, but reaches almost 30\% for the heaviest deuterated isotopologue, leading us to the conclusion that according to the present status of the theory, dissociative recombination is not responsible for the peculiar $^{14}$N/$^{15}$N isotopic ratios of N$_2$H$^+$ observed in the different interstellar molecular clouds. 
\end{abstract}

\maketitle

\section{Introduction}

   N$_2$H$^+$ is one of the first three-atomic charged specimens observed in the interstellar medium (ISM) and one of the key elements of nitrogen's astrochemistry that is less important for terrestrial environments. It 
 provides astronomers with detailed information about the fractional ionization of molecular clouds and the chemistry taking place therein. 

N$_2$H$^+$ has been detected and observed in many different interstellar environments including dark clouds~\citep{turner1974}, translucent clouds~\citep{turner1995}, protostellar cores~\citep{caselli2002}, and photodissociation regions~\citep{fuente1993}. 

In the ISM, N$_2$H$^+$ is mainly formed in proton transfer reactions~\citep{herbst1973,dalgarno1976}
 \begin{equation}
 \mbox{N}_2 + \mbox{H}^+_3 \to \mbox{N}_2\mbox{H}^+ + \mbox{H}_2,
 \label{eq:form1}
 \end{equation}
 and can be destroyed either via proton transfer to abundant molecules, e.g. CO~\citep{mladenovic2014}
 \begin{equation}
 \mbox{N}_2\mbox{H}^+ + \mbox{CO} \to \mbox{HCO}^+ + \mbox{N}_2,
 \label{eq:destr1}
 \end{equation}
 or recombining with slow electrons~\citep{mezei2023}:
  \begin{equation}
 \mbox{N}_2\mbox{H}^+ + e^- \to \begin{cases}
\mbox{N}_2 + \mbox{H} & \\ 
\mbox{NH}  + \mbox{N} & \\
\end{cases}
.
 \label{eq:destr2}
 \end{equation}
 
\noindent N$_2$H$^+$ also plays an important role in the nitrogen-rich planetary atmospheres of solar planets, moons, and exoplanets. For example, in the N$_2$-dominated ionosphere of Titan, N$_2$H$^+$ is formed in ion-atom collisions~\citep{nixon2024}:
 \begin{equation}
  \mbox{N}^+_2 + \mbox{H}_2 \to \mbox{N}_2\mbox{H}^+ + \mbox{H}, 
  \label{eq:form2}
  \end{equation}
  but is quickly lost via electron recombination - eq.~(\ref{eq:destr2}) - or proton transfer reactions~\citep{vuitton2007}
  \begin{equation}
 \mbox{N}_2\mbox{H}^+ + \mbox{CH}_4 \to \mbox{N}_2 + \mbox{CH}^+_5. 
  \label{eq:destr3}
 \end{equation}
 To have a prediction on the abundance of N$_2$ based on N$_2$H$^+$ observations, due to the lack of permanent dipole moment of N$_2$, and thus the lack of observable rotational lines, it is necessary to have a detailed insight into the production and destruction mechanisms of N$_2$H$^+$.

In the recent years, there was a growing interest in the isotopic ratio of interstellar nitrogen ($^{14}$N vs. $^{15}$N) in the view of establishing to which degree the planetary systems inherit their chemical composition from their parent interstellar clouds  \citep[see for example][and references therein]{hily-blant2020}. Moreover, isotopic ratios have long been used to trace the history of cosmic objects such as planets \citep[e.g.]{furi2015, marty2017}. Yet, the current status on the origin of nitrogen in the Solar System still remains elusive~\citep{hily-blant2013, furi2015, hily-blant2020}. One of the main challenges is to determine the sources of isotopic ratio variations of nitrogen in star-forming regions and protoplanetary disks.

Two processes are known that could lead to fractionation -- that is, a deviation of the isotopic ratio in some species, such as HC$^{14}$N/HC$^{15}$N or $^{14}$NH$_3/^{15}$NH$_3$, from the bulk value -- in star-forming regions and protoplanetary disks. One is the isotope-selective photodissociation of N$_2$, which is thought to be the main process in protoplanetary disks \citep{lee2021} but negligible in dense, dark clouds. The second process is known as chemical mass fractionation~\citep{watson1976}: heavier isotopic forms have lower zero-point energies (ZPE). As a consequence of this energy difference, molecules can become enriched in a heavier isotope, through isotope-exchange reactions, when the kinetic temperature of the medium is comparable to, or lower than, the ZPE energy difference, such as in cold dense starless cores. State-of-art measurements of the \foun/\fifn\ ratio in cold clouds in an array of carriers (CN, \ce{HCN}, \ce{HC3N}, \ce{HC5N}, \ce{NH3} and \ce{NH2D}) lead to a consistent value of 330, which was proposed to be the elemental ratio in the present-day solar neighborhood \citep{hilyblant2017, hily-blant2020}, although it could be as low as 270 \citep{adande2012}. These results thus suggest that chemical mass fractionation is at most marginally efficient even in cold clouds. However, accurate determinations of the \foun/\fifn\ ratio in \ce{N2H+} in cold clouds lead to much larger values, from 500 to 1000, which remain unexplained.

\cite{loison2019} and \cite{hily-blant2020} have proposed, that dissociative recombination (DR) of NNH$^+$ with slow electrons might be responsible for isotope-selective depletion measurements of \cite{lawson2011} have shown that, at 300K, the DR of \ce{N2H+} proceeds about $20\%$ faster than that of the heavier $^{15}$N$_2$H$^+$. As modeled by \cite{hily-blant2020}, a larger effect, but in the opposite sense, possible at very low temperatures, could raise the isotopic ratios N$_2$H$^+$/N$^{15}$NH$^+$ and N$_2$H$^+$/$^{15}$NNH$^+$ to values much larger than 330 and up to 1000. Indeed, a factor of two to three enhancement in the DR rate coefficients of the $^{15}$N isotopologues yield values of isotopic ratio for both N$^{15}$NH$^+$ and $^{15}$NNH$^+$ which are consistent with those observed by e.g.~\cite{redaelli2018} \citep[][their Fig. 5]{hily-blant2020}. The enhancement was also found to preserve the slightly lower value of the ratio for N$_2$H$^+/$N$^{15}$NH$^+$ compared to N$_2$H$^+/^{15}$NNH$^+$. However, these effects lack any theoretical or experimental support.

In the present study, our goal is to provide dissociative recombination rate coefficients for a number of N$_2$H$^+$ isotopologues using the normal-mode approximation combined with R-matrix theory and multi channel quantum defect theory (MQDT) as presented in~\cite{slava2018} or~\cite{mezei2019,mezei2023}, with the hope to help to solve the existing isotopic ratio puzzle.
 
The DR of N$_2$H$^+$ has been studied experimentally quite intensively: the Flowing Afterglow Langmuir Probe (FALP) experiments of~\cite{smith1984,poterya2005,lawson2011} resulted in thermal rate coefficients around room temperature. 
  In an experiment using absorption techniques for the final products of~\cite{adams1991}, it was found that the DR of N$_2$H$^+$ predominantly goes into the N$_2+$ H (upper) branch of eq. (3). 
 This was confirmed by a second CRYRING (CRYRING ion ring, Manne Siegbahn Institute of Physics, Stockholm) measurement given by~\cite{vigren2012} - providing DR cross sections up to 10 eV collision energies - saying that more than 90\% of the DR goes into the  N$_2+$ H branch.
 
 Complementary to the experimental studies, extensive theoretical effort has been made regarding the DR of N$_2$H$^+$. \cite{talbi2007,kashinski2012} have reported a series of structure calculations performed at linear geometries with the N-N/N-H bond lengths frozen, identifying and characterizing both possible - experimentally explored - DR pathways, and showing that N$_2$ + H should be favored over N + NH. 
 
A simple model was developed and successfully applied for HCO$^+$ by~\cite{jungen2008} showing that the DR of polyatomic molecular cations may be quite efficient when only the indirect pathway is relevant. 

\cite{douguet2012} have developed a method where the ionic target was described within the normal-mode approximation combined with the Kohn-variational method responsible for the electron scattering calculations, and vibronic frame transformation leading to the S-matrix and consequently to cross sections. The method was applied for the DR of different polyatomic molecular species, see e.g.~\cite{mezei2019} among others for N$_2$H$^+$~\citep{fonseca2014}, and DR cross sections were found that agree well with the CRYRING measurements~\citep{vigren2012}.

In a more recent paper, \cite{mezei2023} have revisited the DR of N$_2$H$^+$. In the framework of a 1D diatomic model using stepwise MQDT calculations based on the molecular structure data with frozen N-N bounds obtained by quantum chemistry calculations of~\cite{talbi2007} and \cite{kashinski2012} we managed to point out that for the linear N$_2$H$^+$ cation, the direct mechanism is negligible. Moreover, performing 3D calculations within the normal mode approximation, combined with the R-matrix theory and MQDT, we obtained a new DR cross section that compares well with the preceding theoretical results and gives slightly better agreement with the storage ring measurements~\citep{vigren2012} for low collision energies.

The present paper is structured as follows: In the section following this Introduction, we describe briefly the methods used in the calculation. This is followed by the obtained results and discussions. The paper ends with conclusions and perspectives.

\section{Theory}

The multi-dimensional nature of the vibrational motion of polyatomic ions makes the theoretical study of their dissociative recombination more difficult than that of the diatomic ones. We explore this complexity with a method based on the normal mode approximation combined with the R-matrix theory~\citep{tennyson2010} and vibronic frame transformation within MQDT. The normal mode approximation is responsible for the vibrational manifold of the target and neutral and usually provides a good description of the vibrational dynamics of molecular systems near the equilibrium geometry. We perform R-matrix calculations to describe the electron scattering on the target. This provides us with the electronic reaction or K-matrix, avoiding the a priori production of the potential energy surfaces (PESs) of the electronic states of the neutral and their electronic couplings with the ionization continuum. And finally, we perform the frame transformation from the body-fixed (or molecular) frame to the laboratory frame.

In the present calculation we follow the general ideas presented by~\cite{mezei2023}, relying on the following assumptions: (1) the rotation of the molecule is neglected, (2) the cross section is averaged over the autoionizing resonances, (3) the autoionization lifetime is assumed to be much longer than the predissociation lifetime, (4) the harmonic approximation is used to describe the vibrational states of the core ion and (5) the different masses of the isotopologues are considered both at the level of electron structure and electron scattering calculations. The electron-induced recombination of the target is described in the same way as its vibrational excitation.

\subsection{Normal modes of the target}
\label{sec:molpro} 

In the equilibrium geometry of the ground electronic state of the linear triatomic N$_2$H$^+$, the 14 electrons in the $C_{2v}$ point group symmetry are distributed over the $(1\sigma)^2(2\sigma)^2(3\sigma)^2(4\sigma)^2(5\sigma)^2(1\pi)^4 $ electronic configuration.  
It has three normal modes: The symmetric $(q_4,\omega_4)$ and asymmetric stretching $(q_3,\omega_3)$ and the doubly degenerate bending $(q_{1,2},\omega_{1,2})$. 

\begin{figure*}
	\includegraphics[width=0.32\textwidth]{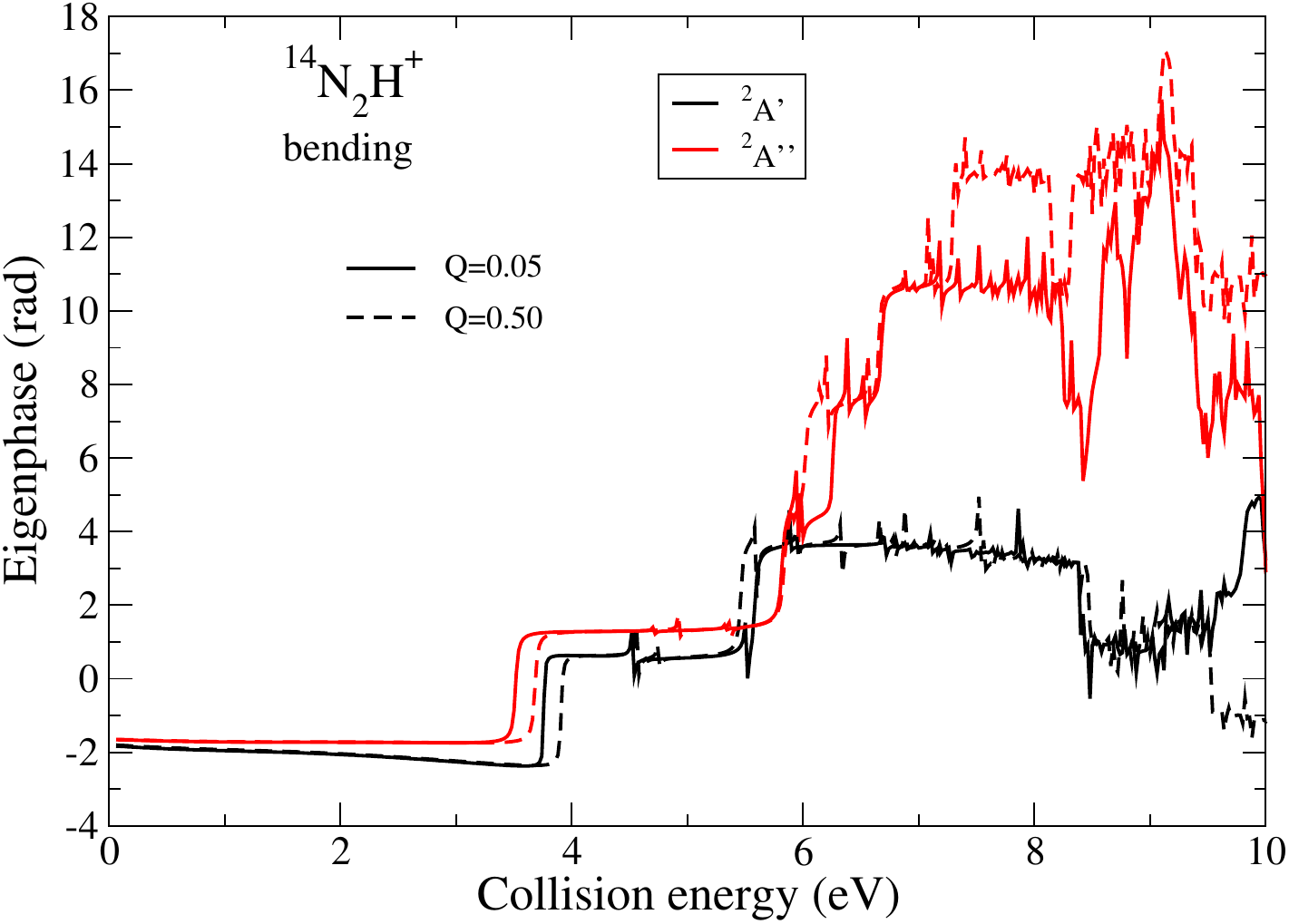}
	\includegraphics[width=0.32\textwidth]{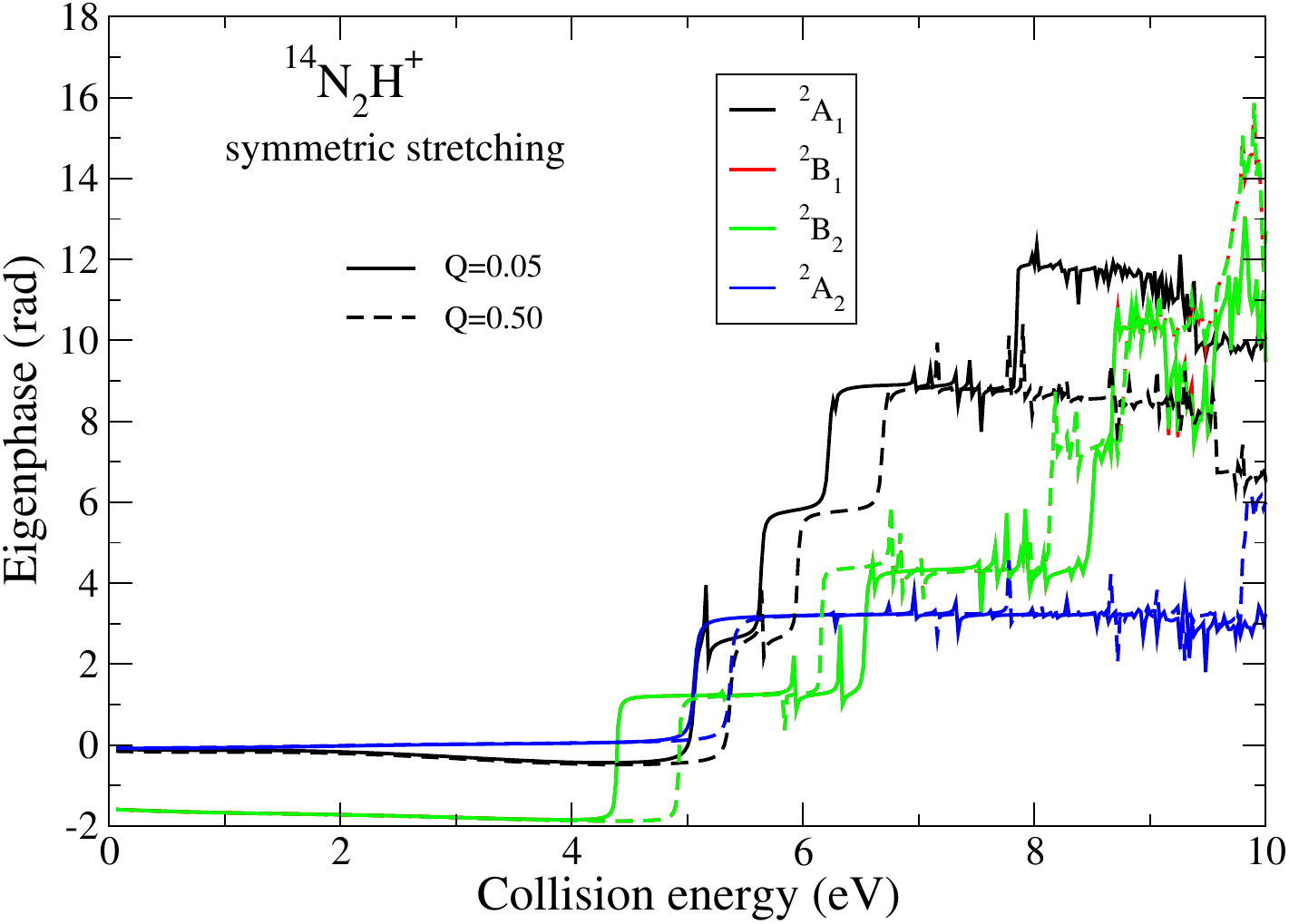}
	\includegraphics[width=0.32\textwidth]{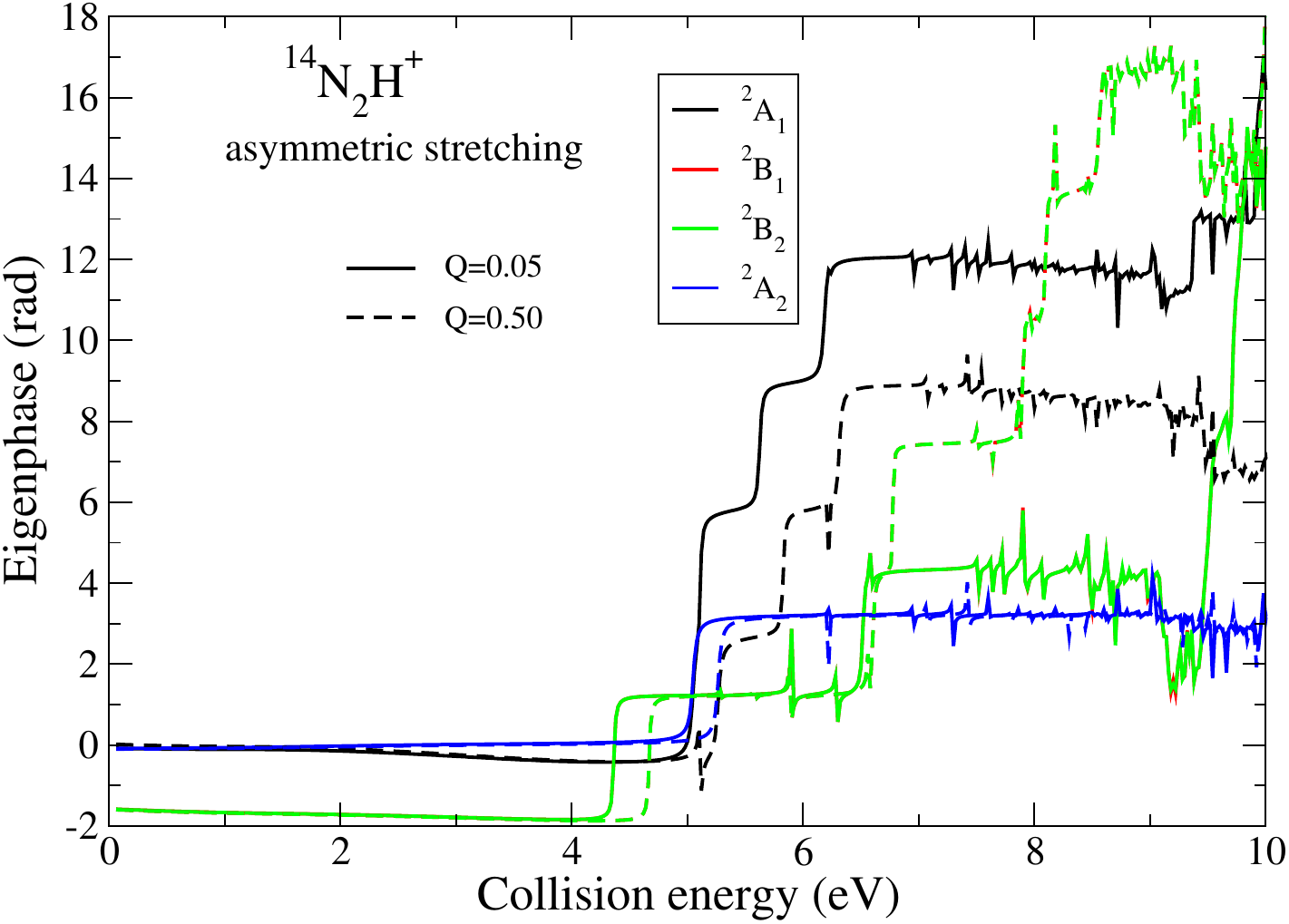}
    \caption{
   The sum of eigenphases for the three different normal modes of $^{14}$N$_2$H$^+$ cation as a function of the electron scattering energy for a displacement of $Q_i = 0.05$ and $0.5$ along each normal mode. The eigenphases for the $^2B_1$ (red) and $^2B_2$ (green) symmetries are on top of each other for the two stretching modes.}
    \label{fig:1}
\end{figure*}

The electron structure calculations were performed using the Molpro program suite~\citep{molpro}, while the level of theory and complete active space (CAS) were chosen considering the existing constraints introduced by the electron scattering calculations. The convergence criteria were explained in more detail in our previous study~\citep{mezei2023}, and the most relevant outputs (the optimized equilibrium geometries, the normal mode frequencies, and dipole moments) obtained for the main ($^{14}$N$_2$H$^+$) isotopologue are summarised in Table~1. of the same study. Analyzing the dependence of ground state energy and permanent dipole moment of N$_2$H$^+$ as the function of standard existing basis functions and the size of the CAS, we have found the cc-pVTZ as a basis function, and 2 frozen, 9 active, and more than 60 external/virtual orbitals as the optimal and most efficient combination. The level of theory was chosen CASSCF.

Considering the above ingredients, we have performed the structure calculations for all 8 relevant isotopologues, which together with their normal mode frequencies and dipole moments are listed in Table~\ref{tab:1}.

\begin{table}
	\centering
	\caption{Characteristics of the linear N$_2$H$^+$ cation and its isotopologues: normal mode frequencies and permanent dipole moments.}
	\label{tab:1}
	\begin{tabular}{lcccc} 
		\hline
		\hline
	isotopologue   &  $\omega_{1,2}$  & $\omega_{3}$ & $\omega_{4}$ & $\mu_e$\\
		  &       (cm$^{-1}$) & (cm$^{-1}$)  & (cm$^{-1}$)  & (debye)\\
		\hline
		\hline
$^{14}$N$_2$H$^+$ & 710.58 & 2274.15 & 3368.87 & 3.419 \\
$^{14}$N$_2$D$^+$	& 562.34 & 2043.17 & 2698.59 & 3.171 \\
$^{14}$N$^{15}$NH$^+$ & 705.80 & 2245.77 & 3352.89 & 3.340 \\
$^{14}$N$^{15}$ND$^+$ & 556.22 & 2036.15 & 2659.44 & 3.103 \\
$^{15}$N$^{14}$NH$^+$	& 709.49 & 2236.37 & 3366.99 & 3.515 \\
$^{15}$N$^{14}$ND$^+$	& 560.90 & 2018.53 & 2682.66 & 3.273 \\
$^{15}$N$_2$H$^+$ &704.66 & 2206.82 & 3351.05 & 3.436 \\
$^{15}$N$_2$D$^+$ & 554.77 & 2010.21 & 2644.19 & 3.203 \\
		\hline
		\hline
	\end{tabular}
\end{table}

\subsection{Fixed-geometry scattering matrix}

The next step, following the characterization of the target, is the electron scattering calculation performed on the “electron+target” system. This part is the weak link in our method, since most of the higher level theories and basis functions are not tractable for the scattering calculations performed with UK R-matrix-based Quantemol-EC (QEC) program suite, see e.g.~\cite{cooper2019,masin2020}. The calculations were performed in the abelian subgroup $C_s$ (bending mode) and $C_{2v}$ (stretching modes) and the target ion was assumed to be in its ground electronic state. We have performed Configuration Interaction Self-Consistent Field (equivalent to the CASSCF) calculations using Hartree-Fock orbitals by freezing 6 electrons in the core $1(a')^2, 2(a')^2, 3(a')^2$ molecular orbitals (MOs) for bending mode, and $1(a_1)^2, 2(a_1)^2, 3(a_1)^2$ for the stretching modes. The remaining 8 electrons are kept free in the active space of the $4a', 5a', 6a', 7a', 1a'', 2a''$  MOs for bending and $4a_1, 5a_1, 6a_1, 1b_1, 2b_1, 1b_2, 2b_2$ MOs for stretching modes, respectively.  According to correlation between $C_s$ and $C_{2v}$ point groups, we have $A' \to A_1+B_2$ and $A''\to A_2+B_1$. The detailed analysis of the MOs has shown that the last orbital of the $C_{2v}$ is double degenerate, leading to an apparent discrepancy. Virtual molecular orbitals have been added to the CAS for the augmentation to the continuum orbitals in the following way: 3 virtual MOs of $A'$ and 1 virtual MO of $A''$ symmetries for the bending mode and 3 $A_1$, 1 $B_1$ and 1 $B_2$ virtual MOs for the stretching ones. A total of nine electronic excited target states are included in our electron scattering calculation. We used an R-matrix sphere with a radius of 14 bohr and in the partial wave expansion, we went up to $l=4$. 

At low collision energies, the fixed-nuclei scattering matrix depends only weakly on energy, which can be best parametrized through the eigenphase sum. The figures given in Fig.~\ref{fig:1} show the obtained eigenphase sums for the three normal modes of the main isotopologue up to 10 eV collision energy. One can observe that the first resonance (responsible for the direct mechanism) appears at 4 eV collision energy, so for collision energies below 4 eV the 3D model based on the indirect mechanism only will describe correctly the dissociative recombination process for N$_2$H$^+$. 

To have the DR cross section, one needs the low-energy scattering matrix calculated in our case using the R-matrix theory via the QEC interface. The calculated electronic $K_{l\lambda,l'\lambda'}(\bf{q})$ reaction matrix for a given geometry is converted into the electronic S-matrix applying a Cayley transform:
\begin{equation}\label{eq:eSmat}
\boldsymbol{\mathcal{S}}=\frac{\boldsymbol{I}+i \boldsymbol{\mathcal{K}}}{\boldsymbol{I}-i \boldsymbol{\mathcal{K}}}
\end{equation}
The vibronic scattering matrix is obtained following the frame transformation of the electronic S-matrix:
\begin{equation}\label{eq:ftsmat}
S_{l\lambda,l'\lambda'}^{\textbf{v},\textbf{v}'}=\int \psi_{\textbf{v}}(\textbf{\textbf{q}})S_{l\lambda,l'\lambda'}(\textbf{q})\psi_{\textbf{v}'}(\textbf{q})d\textbf{q}
\end{equation}
Here $(l\lambda)/(l'\lambda')$ denotes the initial/exit channel of the electron-N$_2$H$^+$ collision system, $l$ being the electron angular momentum and $\lambda$ its projection on the molecular axis. $\psi_{\textbf{v}'}(\textbf{q})$ and $\psi_{\textbf{v}}(\textbf{q} )$ stand for the initial and final vibrational states of the ionic core, respectively, and $\textbf{q}$ denotes collectively the coordinates of all nuclei.
Once the S-matrix is known, the cross sections for the reactive/elastic/inelastic processes are also known.
%
\section{Results and discussions}

\subsection{Cross sections}
Using the harmonic approximation 
\begin{equation}\label{eq:ftsmat}
\psi_{\textbf{v}}(\textbf{q})=\Pi_i^3\chi_{v_i}(q_i),\: \chi_{v_i}(q_i)=\left(\frac{1}{\pi}\right)^{1/4}\frac{1}{\sqrt{2^{v_i}v_i!}}e^{-\left(\frac{q^2_i}{2}\right)}\mathcal{H}_{v_i}(q_i),
\end{equation}
with a restriction to transitions $v_i=0$ to $v_i'=1$ in each mode, 
and expanding the vibronic scattering matrix elements to first order in the normal coordinates, we obtain the following form for the cross section, according to~\cite{slava2018}:
\begin{equation}\label{eq:drxsec}
\sigma^{\text{DR}}(\epsilon)=\frac{\pi \hbar^2}{2\mu\epsilon}\sum_{i=1}^{3}\frac{g_i}{2}\sum_{l\lambda,l'\lambda'}\left| \frac{\partial S_{l\lambda,l'\lambda'}(q_i)}{\partial q_i}\right|^2 \theta\left(\hbar \omega_i-\epsilon\right)
\end{equation}
Above, $\mathcal{H}_{v_i}(q_i)$ is the Hermit polynomial, $\mu$ is the reduced mass of the electron–cation system, $\epsilon$ stands the incident energy of the electron, and $\theta$ is the Heaviside step function. $i$ runs over all three normal modes: two (symmetric and asymmetric) stretching modes with respective frequencies $\omega_3$ and $\omega_4$ and the corresponding (dimensionless) coordinates $q_3$ and $q_4$, and a doubly degenerate bending mode with a lower $\omega_1=\omega_2$ frequency and $q_1$ and $q_2$ coordinates. Finally, $g_i$ $(i=1,2,3)$ is the degeneracy of the modes, being equal to one for the stretching modes ($g_2=g_3=1$) and two for the bending mode ($g_{1}=2$).
It remains to calculate the derivative of the scattering matrix ${\partial S_{l\lambda,l'\lambda'}^{vv'}}/{\partial q_i}$ with respect to the normal coordinate $q_i$, which is numerically evaluated for two $q_i$ values, typically 0.05 and 0.5, keeping the other normal coordinates in the equilibrium geometry ($q_{i'} = 0, i'\ne i$).

\begin{figure}
	\includegraphics[width=0.95\columnwidth]{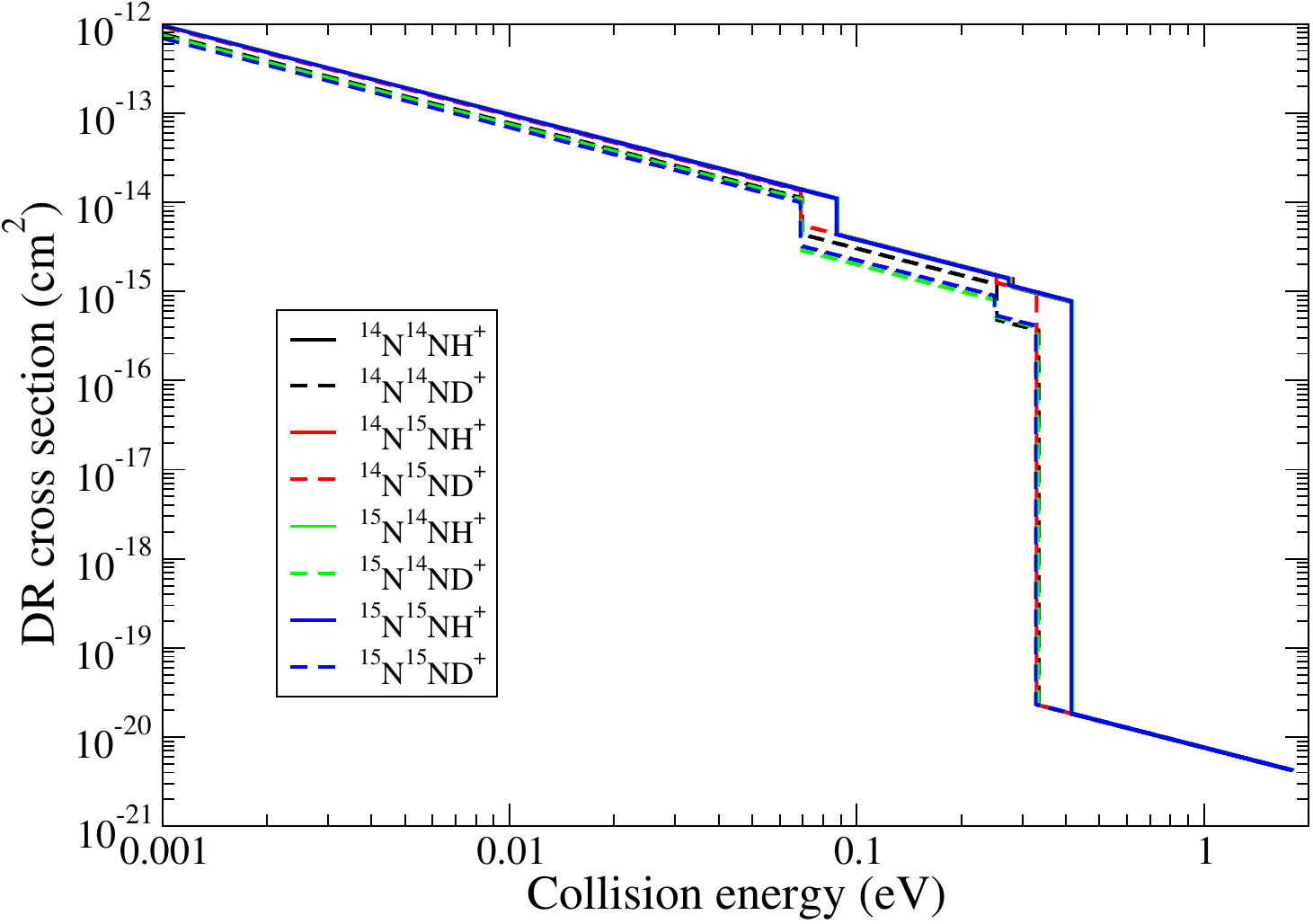}
    \caption{
    Dissociative recombination cross sections for the eight N$_2$H$^+$ isotopologues including $^{14}$N, $^{15}$N, H, and D as the function of the collision energy. The continuous lines stand for the H-bearing isotopologues, while the dashed lines are for the deuterated ones. }
    \label{fig:2}
\end{figure}

Figure~\ref{fig:2} shows the calculated DR cross sections for all 8 isotopologues.  As general rules, one can say, that at very low scattering energies, i.e. below 30 meV, the DR cross section is a smooth function and inversely proportional to the incident energy of the electron, as predicted by the Wigner threshold law, whereas at higher energies it exhibits a drop at each vibrational threshold given in Table~\ref{tab:1}.  The solid lines with different colors given in Fig.~\ref{fig:2} stand for the hydrogen-containing isotopologues, while the dashed lines represent the deuterated ones. One can immediately see that significant differences we have obtained only for the deuterated isotopologues, while those containing hydrogen are essentially on top of each other. This is due to the smaller changes in the reduced mass of the H-bearing isotopologues.

The comparison of the present cross sections to the existing experimental measurements or other theoretical calculations is out of the purpose of the present manuscript\footnote{ We note that measurements on the N$_2$H$^+$ and $^{14}$N$^{15}$NH$^+$ ions are currently in progress at the Cryogenic Storage Ring in Heidelberg (Oldrich Novotny, private communication).}. This was done for the main isotopologue in our previous work~\citep{mezei2023}. In Fig. 5 of that study, we have compared our calculated cross section with the CRYRING measurements published by~\cite{vigren2012} (blue dots) and with the theoretical result of~\cite{santos2014} (green curve). We found that our cross section is in very good agreement with the experimental results at low collision energies, below 30 meV, while at higher collision energies the agreement still remains within a factor of two. Moreover, our results have improved the previous theoretical calculations~\citep{santos2014} for temperatures relevant to the cold ISM.

The present theoretical approach is able to describe the competing inelastic processes like vibrational (de-)excitation (VE or VdE), for the present case only for one quantum in each normal mode of the target. The excitation cross sections for changing two or more quanta in one normal mode are much smaller (due to the propensity rule) and would require extending the harmonic approximation to second or even higher-order terms of the expansion. Since we are focusing on the cold ISM, we neglect the higher excitations. 
The present model suggests that the probability of excitation of the ion by the electron is described by the same physics as dissociative recombination, the excited neutral complex either decays by freeing up an electron and leaving the vibrationally excited ion, or the electron is captured in a Rydberg resonance attached to that particular vibrational state. In other words, if the electron is captured by the ion, but the electron energy is not sufficient to excite the ion and leave it again, the system will most likely dissociate, rather than autoionize.
Thus, the vibrational excitation cross section reads as:
\begin{equation}\label{eq:vexsec}
\sigma_i^{\text{VE}}(\epsilon)=\frac{\pi \hbar^2}{2\mu\epsilon}\frac{g_i}{2}\sum_{l\lambda,l'\lambda'}\left| \frac{\partial S_{l\lambda,l'\lambda'}(q_i)}{\partial q_i}\right|^2 \theta\left(\epsilon - \hbar \omega_i\right)
\end{equation}
 
\subsection{Rate coefficients}

\begin{figure}
	\includegraphics[width=0.95\columnwidth]{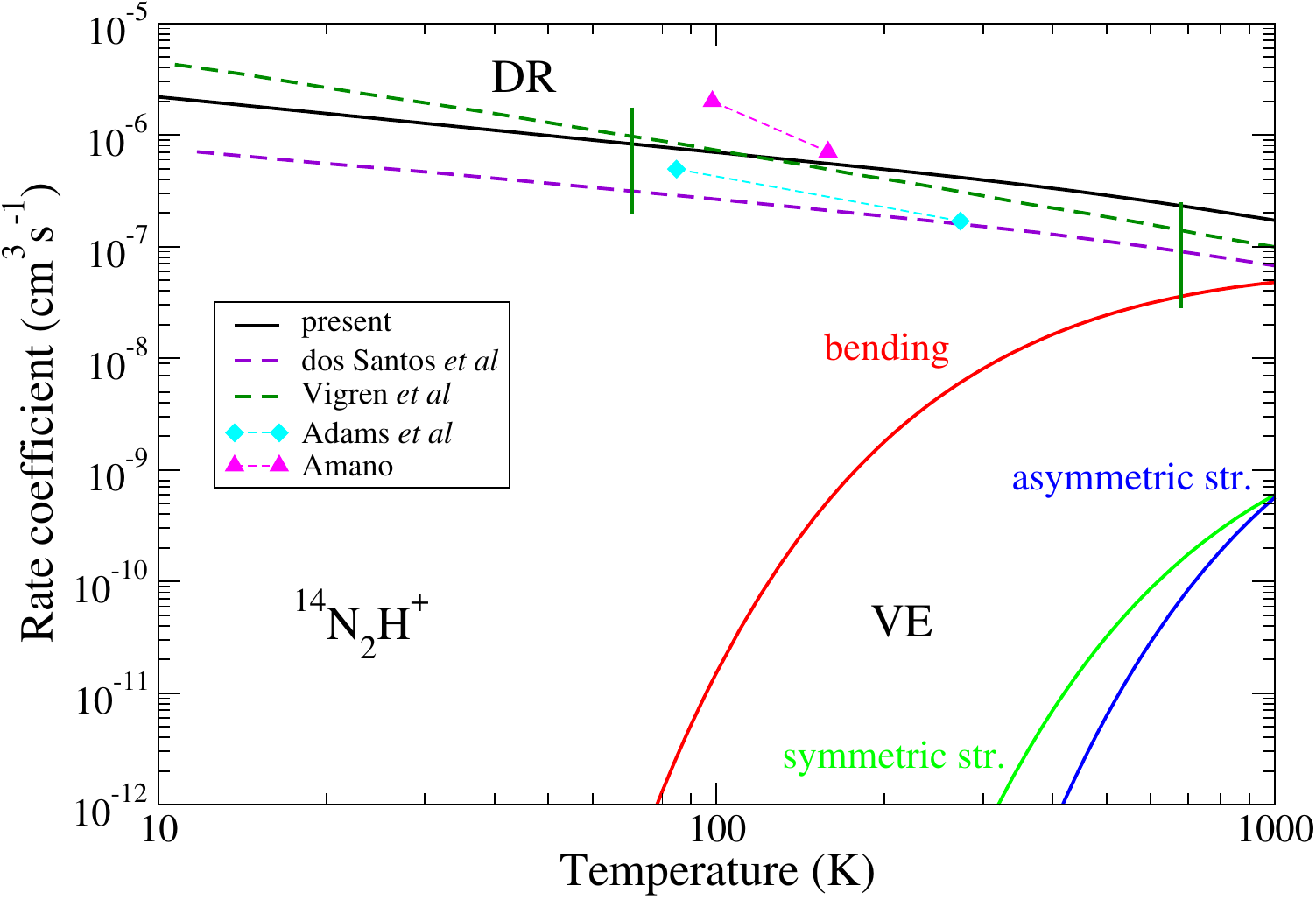}
    \caption{
    Dissociative recombination (black) and vibrational excitation (red, green, and blue) rate coefficients for the $^{14}$N$_2$H$^+$ cation in collision with slow electrons as a function of the electronic temperature. Our DR results are compared with three experimental - dark green dashed line with error bars~\citep{vigren2012}, cyan dashed line with diamonds~\citep{adams1984} and magenta dashed line with triangles~\citep{amano1990} - and one theoretical~\citep{santos2014} results.}
    \label{fig:3}
\end{figure}

The thermal rate coefficients are determined by convoluting the cross section with the Maxwell energy distribution function of the free electrons:
\begin{equation}\label{eq:rate}
\alpha(T)=\frac{2}{kT}\sqrt{\frac{2}{\pi mkT}}\int_0^{\infty}\sigma(\epsilon)\epsilon \exp(-\epsilon/kT)d\epsilon,
\end{equation}
where $m$ and $\epsilon$ are the electron's mass and collision energy, $T$ is the temperature and $k$ the Boltzmann constant. 

Figure~\ref{fig:3} shows the DR and VE ($\Delta v=1$) rate coefficients calculated for the main isotopologue. We compare our calculated DR rate coefficients with those measured in the CRYRING storage ring by~\cite{vigren2012} and the FALP experiments done by~\cite{adams1984} and by~\cite{amano1990}, as well as with the rates calculated by~\cite{santos2014}. Our rate coefficients compare better with the experiments than the other theory, agreeing within a factor of two with the storage ring results. Moreover, we have calculated the $0 \to 1$ vibrational excitation rate coefficients for the three normal modes shown in red (bending), green (symmetric stretching), and blue (asymmetric stretching) solid lines, emphasizing the importance of the bending mode over the two stretching modes. 

\begin{figure}
	\includegraphics[width=0.95\columnwidth]{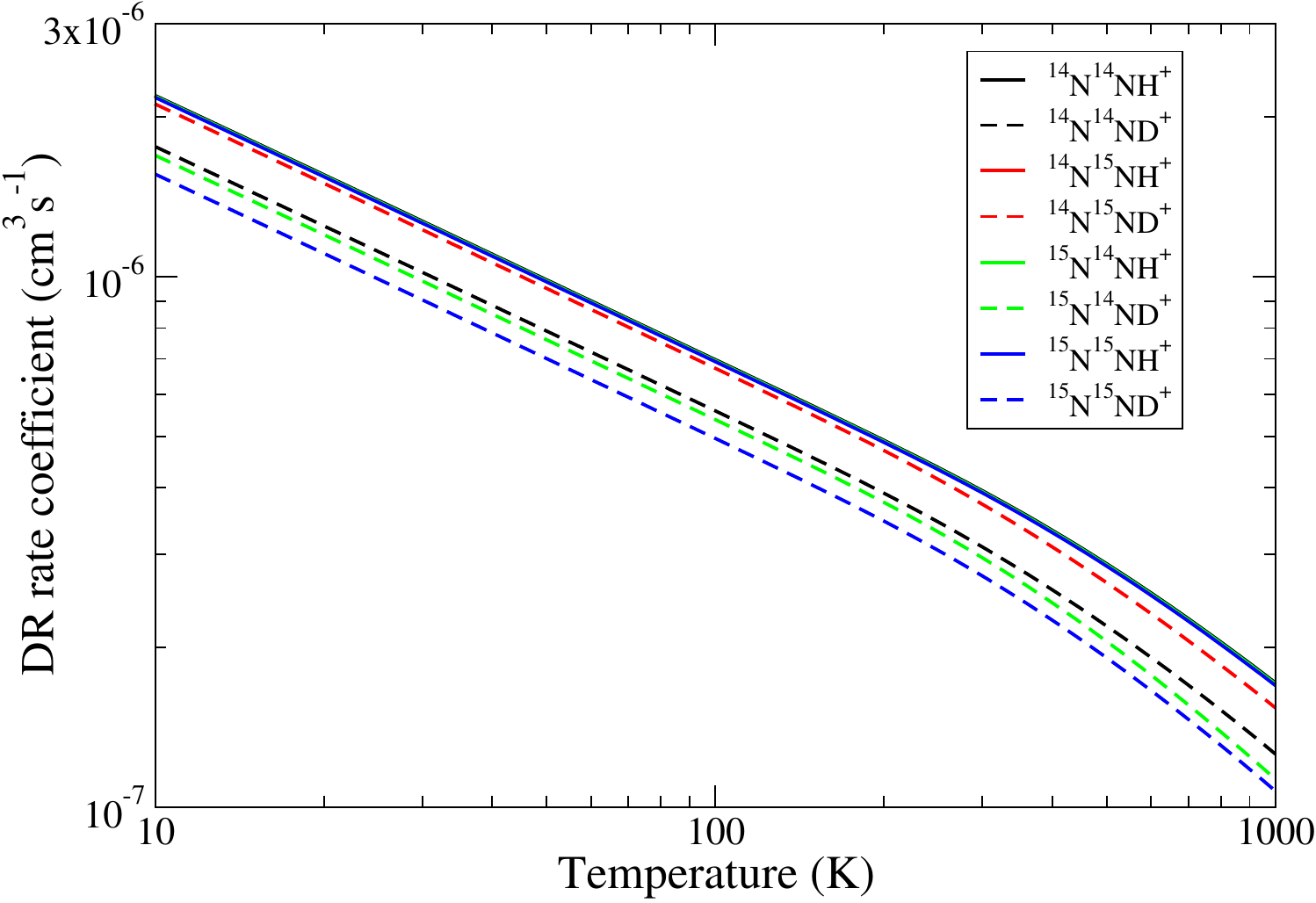}
    \caption{
    Dissociative recombination rate coefficients for the N$_2$H$^+$ and isotopologues. We have used the same color code as in figure~\ref{fig:2}.}
    \label{fig:4}
\end{figure}

\begin{figure}
	\includegraphics[width=0.95\columnwidth]{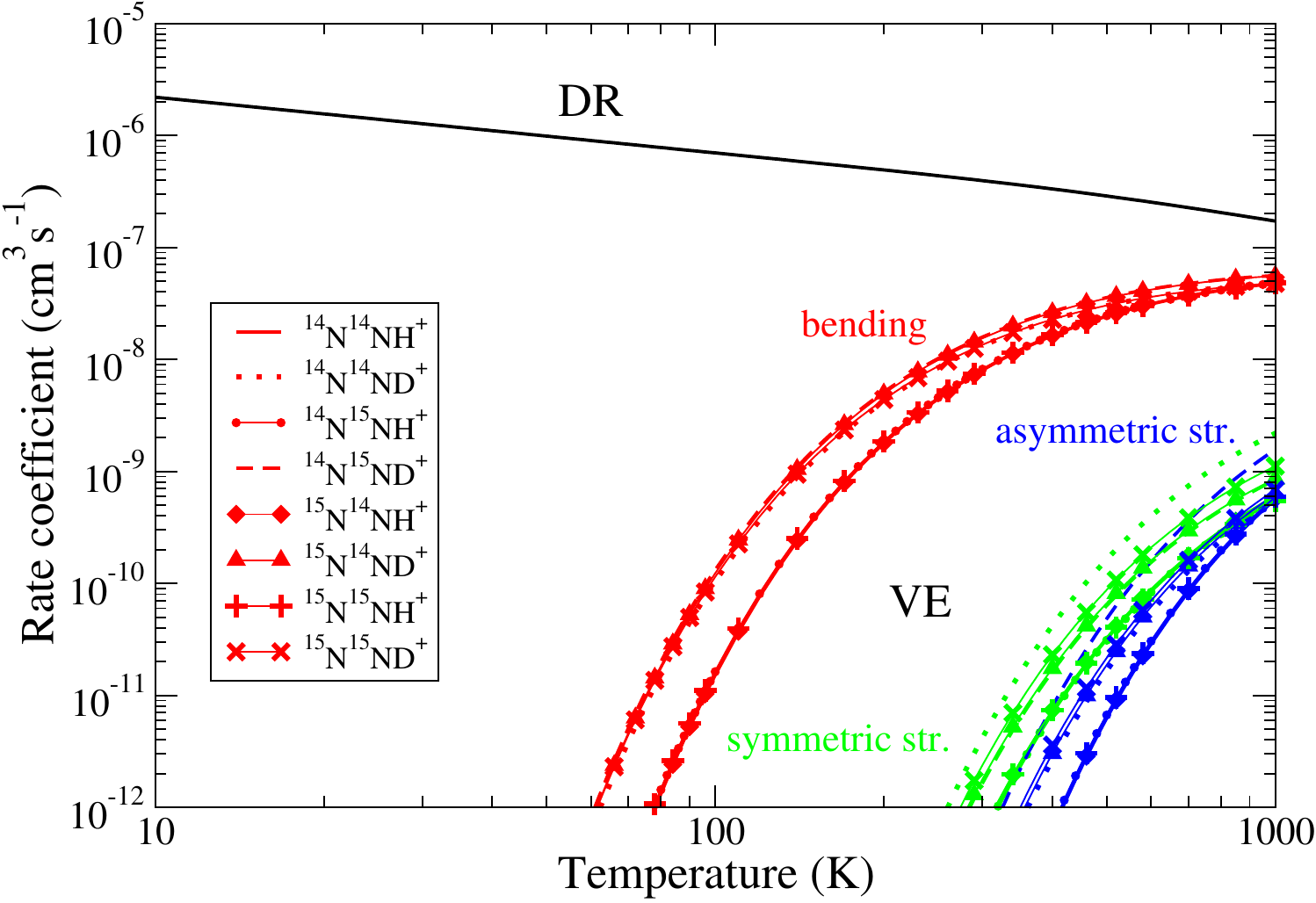}
    \caption{
    $0\to 1$ vibrational excitation rate coefficients for the N$_2$H$^+$ and isotopologues for all three normal modes. DR rate coefficients for the main isotopologue are also given for comparison.}
    \label{fig:5}
\end{figure}

In Figures~\ref{fig:4} and \ref{fig:5} we show the main results of the present study, namely the isotopic effects obtained for the DR and VE processes. Using the color code introduced in Fig.~\ref{fig:2}, in Fig.~\ref{fig:4} we present DR rate coefficients for all 8 isotopologues. One can see that the change in the mass of the nitrogen has no considerable effect on the rate coefficients of the hydrogen-containing isotopologues, the DR rate coefficients are essentially on top of each other. This is consistent with the minimal change of the reduced mass of the target when changing the masses of the nitrogen atoms. On the contrary, the changes in the atomic masses of the nitrogen have a more considerable impact on the rate coefficients of the deuterated targets. Not only the normal mode frequencies and the reduced mass of the target can influence the rate coefficients of the isotopologues, but the position of the mass center of the nitrogen subsystem relative to the center of mass of the target too, leading to larger differences that can be seen in the figure. The relative differences to the main isotopologue are given in Table~\ref{tab:2} for three temperature values and compared for two temperatures with the experimental results obtained by~\cite{lawson2011}. We have relative differences below 1$\%$ for the hydrogen bearing targets and about 30$\%$ for the heaviest deuterated isotopologue. One can also see, the satisfactory agreement we obtained with the experiments, except for the hydrogen-bearing isotopologues.

\begin{table}
	\centering
	\caption{Relative differences ($\epsilon$) of the DR rate coefficients for the different isotopologues ($k_{j} (T)$) respective to the main isotopologue ($k_{0} (T)$). Our results are compared with the experimental measurements of~\cite{lawson2011}.}
	\label{tab:2}
	\begin{tabular}{lccccc} 
		\hline\hline
	   &\multicolumn{5}{c}{$\epsilon (T)=\left[\frac{k_{j} (T)}{k_{0} (T)}-1\right]$ ($\%$)}\\[5pt]
 \cmidrule{2-6}       
 &       \multicolumn{3}{c}{present} & \multicolumn{2}{c}{Lawson et al.}\\[3pt]
\cmidrule(lr){2-4} \cmidrule(lr){5-6}		  
&     100 K & 300 K  & 500 K & 300 K  & 500 K \\
		\hline
		\hline
$^{14}$N$^{15}$NH$^+$ & -0.45 & -0.49 & -0.56 & & \\
$^{15}$N$^{14}$NH$^+$	& -0.41 & -0.43 & -0.45 & & \\
$^{15}$N$_2$H$^+$ & -0.85 & -0.91 & -1.01 & -16.7 & +3 \\
$^{14}$N$_2$D$^+$	& -19.9 & -21.6 & -23.7 & - 23.1 & -17.4 \\
$^{14}$N$^{15}$ND$^+$ & -3.65 & -5.69 & -7.91 & & \\
$^{15}$N$^{14}$ND$^+$	 & -22.9 & -25.2 & -28.6 & & \\
$^{15}$N$_2$D$^+$ & -28.9 & -30.9 & -33.5 & -28.9 & -17.6 \\
		\hline
		\hline
	\end{tabular}
\end{table}

In Figure~\ref{fig:5}, similarly to DR, we present the rate coefficients for the $0\to 1$ VE of all isotopologues. In the figure, as for the guide of the eye, we show the DR rate coefficient of the main isotopologue. In the case of excitation, we can see two bunches of rate coefficients belonging to the hydrogen and deuterium-containing targets for all three normal modes. The difference between the two branches is larger for low temperatures and disappears above 1000 K. The mass dependence of the DR and VE through their initial and final states are very different, and this can be seen in how the rate coefficients vary with the temperature. While for VE the isotopic effects are diminishing as we go towards higher temperatures, in the case of DR we see slightly increasing differences.

\subsection{Chemical models of collapsing clouds}
\newcommand \msol{\ensuremath{{M_{\odot}}}}
\newcommand \ccc{\ensuremath{\text{cm}^{-3}}}

\begin{figure}
    \centering
    \includegraphics[width=0.95\linewidth]{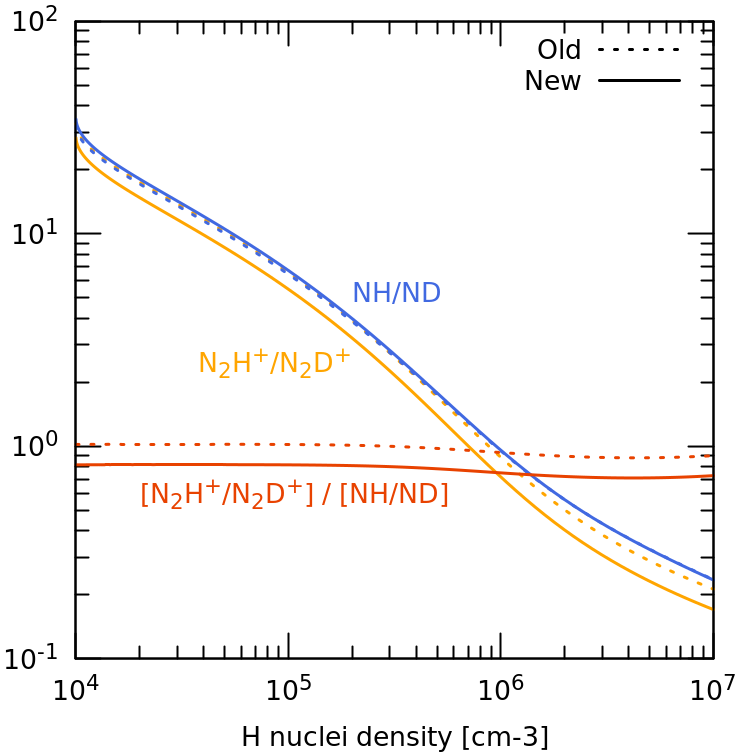}
    \caption{Evolution of the \ce{N2H+}/\ce{N2D+} and NH/ND abundance ratios during the gravitational collapse of an isothermal starless core. The ratio of the two ratios is also shown in red. The cloud undergoes a Larson-Penston-type collapse. See \cite{hily-blant2020} for details.}
    \label{fig:collapse}
\end{figure}

In this section, a chemical model of a collapsing pre-stellar core is combined with the (updated) University of Grenoble Alpes Astrochemical Network (UGAN) \citep{hilyblant2018}. While the small relative differences (below 1\%) reported above for the hydrogen bearing isotopologues of N$_2$H$^+$ can be safely neglected, it is attractive to test the impact of a decrease, by 20\%, of the DR rate of \ce{N2D+} compared to \ce{N2H+}, on the chemical abundances of these two isotopologues, and their daughter species through DR, namely ND and NH. Figure \ref{fig:collapse} shows the results for a 7~\msol\ cloud with an initial H nuclei density of 10$^4$~\ccc\ and a 10~K kinetic temperature. The initial chemical conditions correspond to the steady-state abundances at 10~K and a constant 10$^4$~\ccc\ density, in which the gas is essentially molecular. The calculation follows a fluid particle as the collapse proceeds, which thus sees its density increasing with time \citep[see][for more details]{hily-blant2020}. The results show the \ce{N2H+}/\ce{N2D+} and NH/ND abundance ratios as a function of the H nuclei density for two sets of calculations, one with equal DR rates (dashed lines labeled \textit{Old}) and one with a \ce{N2D+} DR rate that is 20\% smaller (full lines labeled \textit{New}). When the DR rates are equal, both ratios are very close to each other: this is why the \ce{N2H+}/\ce{N2D+} ratio should be a very good proxy for the NH/ND ratio which is hard to quantify in dense clouds \citep{bacmann2016}. With the revised DR rate, the NH/ND ratio is only very marginally affected while \ce{N2H+}/\ce{N2D+} is reduced by $\sim 20\%$. Thus, this ratio remains a good proxy for the NH/ND ratio, although a $\sim 20\%$ correction factor must now be applied.

\section{Conclusions}
In the present work, we have studied the isotopic effects in the dissociative recombination and vibrational excitation of the astrochemically relevant N$_2$H$^+$ molecular cation, by changing the masses of all atomic ingredients.

By using a 3D model calculation in the framework of the normal mode approximation combined with R-matrix theory and Multichannel Quantum Defect Theory we have determined the DR and VE cross sections and thermal rate coefficients for 8 isotopologues by considering all three normal modes: the doubly degenerate bending and symmetric and asymmetric stretching.
Our DR rate coefficients for N$_2$H$^+$ agree well with the storage-ring measurements and, at very low energies, slightly improve the results of the previous theoretical calculations for temperatures relevant to the cold interstellar media. Our calculations show that the relative differences respective to the main isotopologue are below 1$\%$ for the hydrogen-containing isotopologues, and reach about 30$\%$ for the heaviest deuterated isotopologue ($^{15}$N$_2$D$^+$).

The present study concludes that dissociative recombination is not responsible for the peculiar isotopic ratios (N$_2$H$^+$/$^{15}$NNH$^+$ and N$_2$H$^+$/N$^{15}$NH$^+$) observed in the interstellar medium, suggesting that crucial fractionation processes, in the gas or on the surface of the icy grains, are still missing in the current astrochemical networks. 
%
\begin{acknowledgements}
The authors acknowledge support provided by the R\'egion Normandie, FEDER, and LabEx EMC3 through projects PTOLEMEE, COMUE Normandie Universit\'e, the Institute for Energy, Propulsion and Environment (FR-IEPE), as well as the European Union through COST action MD-GAS (CA18212).  The authors acknowledge the contribution of the Agence Nationale de la Recherche (ANR) through the MONA project. This work has received support from the Programme National "Physique et Chimie du Milieu Interstellaire" (PCMI) of CNRS/INSU with INC/INP co-funded by CEA and CNES. JZsM, AO and DS are grateful for financial support from the National Research, Development and Innovation Fund of Hungary, under the  FK 19 funding schemes with Project Number FK 132989. 
\end{acknowledgements}
%
\section*{Data Availability}
The data underlying this article will be shared on reasonable request to the corresponding author.

%
   \bibliographystyle{apsrev4-1} 
   \bibliography{n2hpiso} 
%

\end{document}